\documentclass{elsart}
\usepackage{fleqn}
\usepackage{graphicx}
\usepackage{amsmath}
\usepackage{amsfonts}
\usepackage{amssymb}
\def\be{\begin{equation}}
\def\ee{\end{equation}}
\def\bea{\begin{eqnarray}}
\def\eea{\end{eqnarray}}
\def\ep{\epsilon}

\def\TT{{\mathcal T}}
\def\NN{{\mathcal V}}

\begin{document}

\begin{frontmatter}

\title
{Simple solutions of relativistic hydrodynamics\\ 
for systems with ellipsoidal symmetry} 

\author[KFKI,USP]{T. Cs\"org\H o,\thanksref{tamas}} 
\author[UIB,KFKI]{L. P. Csernai\thanksref{laci}} 
\author[USP]{Y. Hama,\thanksref{yogiro}} 
\author[UFRJ]{T.  Kodama\thanksref{takeshi}} 
\address[KFKI]{MTA KFKI RMKI, H - 1525 Budapest 114, POBox 49, 
Hungary} 
\address[UIB]{Section for Theoretical Physics, Department of Physics,\\
	University of Bergen, Allegaten 55, 5007 Bergen, Norway}
\address[USP]{IF - USP, C.P. 66318, 05389-970 S\~{a}o Paulo - SP, 
Brazil}
\address[UFRJ]{IF - UFRJ, C.P. 68528, 21945-970 Rio de Janeiro - RJ, Brazil}
\thanks[tamas]{Email: csorgo@sunserv.kfki.hu}
\thanks[laci]{\phantom{Email:} csernai@fi.uib.no}
\thanks[yogiro]{\phantom{Email:} hama@fma.if.usp.br}
\thanks[takeshi]{\phantom{Email:} tkodama@if.ufrj.br}

\date{May 26, 2003}

\begin{abstract} 
Simple, self-similar, analytic solutions of (1+3)-dimensional 
relativistic hydrodynamics are presented for ellipsoidally 
symmetric finite fireballs corresponding to non-central collisions 
of heavy ions at relativistic bombarding energies. The hydrodynamical
solutions are obtained for a new, general family of equations of state
with the possibility of describing phase transitions. 
\end{abstract}

\end{frontmatter}

\maketitle

\section{Introduction} 
Recently, new families of exact analytic solutions of relativistic 
hydrodynamics have been found in a Hungarian-Brazilian collaboration. 
The simplest case corresponds to the (1+1)-dimensional expansions 
\cite{s1}: it consists of a class of 1-dimensional scaling flows 
with a proper-time dependent pressure, but characterized with an 
arbitrary rapidity distribution and a temperature field that is 
coupled to the rapidity distribution, thus overcoming one of the 
shortcomings of the well-known Hwa-Bjorken solution \cite{Bjorken}, 
which has a flat rapidity distribution. Physically, the situation 
described in \cite{s1} may characterize e.g. soft collisions of 
hadrons at high bombarding energies. These solutions are obtained 
for a broad class of equations of state that may include massive 
particles and an arbitrary constant of proportionality between the 
kinetic energy density and the pressure. The (1+1)-dimensional 
solutions have been generalized to the case of cylindrical symmetry 
in ref.~\cite{sc}, describing a physical situation that may 
correspond to central collisions of heavy ions at ultra-relativistic 
colliding energies, thus, overcoming another shortcoming of Hwa-Bjorken 
solution, which contains no transverse flow. These exact analytic 
solutions, reported in refs.~\cite{s1,sc}, are associated to each 
non-negative scaling function $\NN(s)$ that satisfies the 
normalization condition $\NN(0)=1$, where the argument $s$ is a 
scaling variable that guarantees the self-similarity of the 
solutions. Here we present a broader class of solutions belonging to 
this new family of exact solutions. In particular, we shall describe 
two classes of ellipsoidally symmetric solutions of relativistic 
hydrodynamics, that may be relevant for the study of non-central 
collisions of heavy ions at relativistic bombarding energies. 

In general, exact analytic resolution of 3 dimensional relativistic 
hydrodynamics is a difficult task due to the highly non-linear nature 
of these equations. The complications are sometimes simplified by 
assuming idealized boundary conditions and simplified equations of 
state (e.g. infinite bombarding energy and a massless relativistic 
gas in the case of Hwa-Bjorken's solution~\cite{csernai}). A more 
realistic but analytically more complicated solution had been found 
by Khalatnikov~\cite{Khalatnikov}, following Landau's basic 
ideas~\cite{Landau}, that gave rise to the hydrodynamical approach in 
high-energy physics. Both of these solutions are frequently utilized 
as the basis for estimations of various observables in 
ultra-relativistic nucleus-nucleus collisions~\cite{csernai}. 
 
Analytic solutions of relativistic hydrodynamics were also reported 
recently in refs.~\cite{biro1,biro2}. However, these solutions are 
valid only at the softest point of the equation of state. For more 
realistic situations, the (1+3)-dimensional relativistic 
hydrodynamical equations are frequently solved using various 
numerical methods, for example, recent solutions were obtained with 
the help of smoothed-particle hydrodynamics in 
refs.~\cite{SPH-jpg,SPH-qm01}. 

The exact analytic solutions, reported here, are generalizations of 
the results of refs.~\cite{s1,sc} to the case of 3-dimensional 
relativistic expansions with less and less symmetry. However, we 
emphasize that these results also correspond to generalizations of 
earlier analytic solutions of non-relativistic hydrodynamics. The 
first solutions of this kind have been found by Zim\'anyi, Bondorf 
and Garpman (ZBG) in 1978~\cite{jnr}. The key aspect of the ZBG 
solution is its self-similarity, this property is kept in all the 
subsequent generalizations including cylindrical or ellipsoidal 
symmetries and relativistic flows. The ZBG solution has been 
generalized to the case of spheroidally symmetric expansions in 
ref.~\cite{jde}, to spherically symmetric Gaussian expansions 
in ref.~\cite{cspeter}. From the point of view of the relativistic
generalizations, an important stepping stone was made in 
ref.~\cite{cssol} when a scaling function has been introduced for the 
spherically symmetric non-relativistic solutions of hydrodynamics. 
This was the first case when a whole new family of solutions has been 
found, that assigned an independent solution for every 
(differentiable and integrable) scaling function that satisfied the requirement of positivity ${\TT}(s)>0$ and normalized as 
${\TT}(0)=1$. The ZBG and the spherical Gaussian solutions appeared 
as special choices for the functional form of the scaling function 
${\TT}(s)$. The Gaussian family of {\it spherical} solutions has been 
generalized to Gaussian {\it ellipsoidal} expansions in 
refs.~\cite{ellsol,ellsp}. Recently, the whole family of self-similar 
ellipsoidally symmetric non-relativisitic solutions of hydrodynamics 
has been found in ref.~\cite{csel}. In the present paper, we report 
on the generalization of this family of solutions to the case of 
relativistic expansions with ellipsoidal symmetry and introduce new 
families of relativistic solutions with even less than ellipsoidal 
symmetry. All the solutions reported here have non-trivial 
non-relativistic limiting behaviour.

\section{The equations of relativistic hydrodynamics.} 
Let us adopt the following notational conventions: 
the coordinates are $x^\mu = (t, {\mathbf r}) = (t, r_x, r_y, r_z)$,
$x_\mu = (t, -r_x, -r_y, -r_z)$ and the metric tensor is 
$g^{\mu\nu} = g_{\mu\nu} = \mbox{\rm diag}(1,-1,-1,-1)$. 
The relativistic continuity and energy-momentum-conservation 
equations are 
\begin{eqnarray}
\partial_\mu(n u^\mu) & = & 0\,, \label{e:cont} \\ 
\partial_\nu T^{\mu \nu} & = & 0\,,  \label{e:tnm} 
\end{eqnarray} 
where $n \equiv n(x)$ is the conserved number density, 
$u^\mu \equiv u^\mu(x) = \gamma(1,{\mathbf v})$ is the four-velocity, 
with $u^\mu u_\mu = 1$, and 
$T^{\mu \nu}$ is the energy-momentum tensor. We assume perfect fluid, 
\begin{equation}  
T^{\mu\nu} = (\ep + p) u^\mu u^\nu - p g^{\mu \nu},
\end{equation} 
where $\ep$ stands for the relativistic energy density and $p$ is 
the pressure.

From general thermodynamical considerations and the perfectness of the
fluid one can show that the expansion is adiabatic,
\be
\partial_\mu(\sigma u^\mu)  =  0\,, \label{e:adiab} 
\ee
where $\sigma$ stands for the entrophy density.

For the equations of state, we assume a gas that may contain massive 
conserved quanta, 
\bea
\ep & =& m n + \kappa p\,, \label{e:eos1}\\
p  & = & n T. \label{e:eos2}
\eea 
These equations have two free parameters, $m$ and $\kappa$. 
Non-relativistic hydrodynamics of ideal gases corresponds to the 
limiting case of ${\mathbf v}^2 \ll 1$ and $\kappa = \frac{3}{2}\,$. 
Our solutions, presented below, exist for all values of $m\ge 0$ and 
$\kappa>0\,$. 

Using the continuity equation~(\ref{e:cont})
and the equations of state~(\ref{e:eos1},\ref{e:eos2}),
the  energy-momentum-conservation equations, 
(\ref{e:tnm}) can be transformed to the 
Euler and temperature equations, 
\bea
u_\nu u^\mu \partial_\mu p + (\ep + p)u^\mu \partial_\mu u_\nu - 
\partial_\nu p & = & 0\,, \label{e:rEu} \\
u^\mu \partial_\mu T + \frac{1}{\kappa} T \partial_\mu u^\mu = 0\,. \label{e:rT}
\eea
Expressing the energy density and the pressure in the 3 independent
components of the relativistic Euler equation in terms of $n$ and $T$
with the help of the equations of state, (\ref{e:eos1},\ref{e:eos2}),
one obtains a closed system
of 5 equations (the continuity, the Euler and the temperature equations)
 in terms of 5 variables, $n$, $T$ and ${\mathbf v} = (v_x,v_y,v_z)$. 
 
\section{Self-similar, ellipsoidally symmetric ansatz}
We search for self-similar solutions in which the isotherms at each 
instant are ellipsoidal surfaces, where the number density is 
also constant. We do not discuss here how the major axes of these 
ellipsoids may be rotated in the frame of observation. We describe 
the solution in the natural System of Ellipsoidal Expansion 
(SEE)~\cite{ellsol,ellsp}, where the coordinate axes point to the 
pricipal directions of the expansion. Let us now define the 
following scaling variable 
\begin{equation} 
s=\frac{r_x^2}{X^2} +\frac{r_y^2}{Y^2} + \frac{r_z^2}{Z^2}\,, \label{e:sdef} 
\end{equation} 
where the scale parameters $X = X(t)$, $Y = Y(t)$ and $Z = Z(t)$ 
are assumed to depend only on the time variable $t$. The condition 
$s = const.$, at each instant $t$, defines the isotherms mentioned 
above. As these surfaces are ellipsoids, the solutions belonging to 
this class will be characterized by ellipsoidal symmetry. We assume a 
self-similar expansion of Hubble type, with possibly different Hubble 
constants in all the pricipal directions of the expansion: 
\bea 
v_x(t,{\mathbf r}) & = & \frac{\dot{X}}{X} r_x,  \label{e:vx} \\ 
v_y(t,{\mathbf r}) & = & \frac{\dot{Y}}{Y} r_y,  \label{e:vy} \\ 
v_z(t,{\mathbf r}) & = & \frac{\dot{Z}}{Z} r_z.  \label{e:vz} 
\eea 
where $\dot A = dA(t)/dt$ stands for the derivative of the scale 
parameter $A = \left\{X,Y,Z\right\}$ with respect to time $t$. 

As a straightforward generalization of the results derived in 
refs~\cite{s1,sc}, we find the following new family of ellipsoidally 
symmetric, exact analytic solutions of relativistic hydrodynamics: 
\bea
 s & = & \frac{r_x^2}{\dot X^2_0 t^2} 
       + \frac{r_y^2}{\dot Y^2_0 t^2} 
       + \frac{r_z^2}{\dot Z^2_0 t^2}\,, 	\label{e:ssol}\\ 
 {\mathbf v} & = & \frac{\mathbf r}{t}\hspace{2.cm}{\rm or}  
       \hspace{2.cm}u^\mu=\frac{x^\mu}{\tau}\,, \\ 
 n(t,{\mathbf r}) & = & n_0\left(\frac{\tau_0}{\tau}\right)^3\NN(s), 
  \label{e:nsol} \\ 
 p(t,{\mathbf r})&=&p_0\left(\frac{\tau_0}{\tau}\right)^{3+3/\kappa},  
  \\ 
 T(t, {\mathbf r}) & = & T_0 \left(\frac{\tau_0}{\tau} 
  \right)^{3/\kappa} \frac{1}{\NN (s)}, \label{e:tsol} 
\eea
where $\tau=\sqrt{t^2-{\mathbf r}^2}\,$, $p_0 = n_0 T_0\,$, 
$\dot X_0$, $\dot Y_0$, $\dot Z_0$ are constants and $\NN(s)$ is an 
arbitrary positive, differentiable function of the scaling variable 
$s$ normalized such that $\NN(0) = 1$. The temperature distribution 
also depends on the scaling variable $s$ through a scaling function 
$\TT(s)\,$, which happens to be $\TT(s) = 1/ \NN(s)\,$. Note that 
the pressure $p$ depends only on $\tau$.    

An interesting property of this solution is that the flow and the
pressure fields are spherically symmetric, however, the other thermodynamical
quantities such as the temperature or density distributions are 
characterized by ellipsoidal symmetry.

The lack of acceleration is reflected by the equations 
\bea
	X & = & \dot X_0 t\, , \label{e:xsol}\\ 
	Y & = & \dot Y_0 t\, , \label{e:ysol}\\ 
	Z & = & \dot Z_0 t\, . \label{e:zsol} 
\eea
In this family of new solutions, the flow is three-dimensional, 
accelerationless Hubble-type, $u^\mu = x^\mu/\tau$, which gives 
\be
u^\mu\partial_\mu u_\nu =0. \label{e:noacc}
\ee 
This property, together with the 
$s$-independence of pressure fields, guarantees that the Euler equation be satisfied regardless of the mass $m$ and the value of 
$\kappa$. We have found new, self-similar, scaling solutions to the 
continuity and the temperature equations with ellipsoidal symmetry. 
Thus a new hydrodynamical solution is assigned to each scaling 
function $\NN(s) = 1/ \TT(s)$, similarly to the cases of the 
ellipsoidally symmetric, non-relativistic solutions of 
ref.~\cite{csel}. 

\section{Phase transitions and freeze-out}
The simplest case of the previously discussed new family
of exact solutions of relativistic hydrodynamics  is given by the
choice of  $\NN(s) = 1$. In this case, the temperature
is constant on a hypersurface characterized by a constant proper-time,
and all the other thermodynamical parameters are constants on these surfaces
as well. Thus these parameters depend on $\tau$ only, similarly 
to the Hwa-Bjorken solution~\cite{Bjorken}.

It is possible to generalize the equations of state,~(\ref{e:eos1},\ref{e:eos2})to characterize a new phase of matter and then describe the rehadronization
of a  ``Quark Matter" (QM) state 
to a Hadron gas (H) in a model similar to that of
Gyulassy and Matsui~\cite{gyu-matsui}. 
For simplicity, from now on we refer to the the equations
of state ~(\ref{e:eos1},\ref{e:eos2}), describing massive, conserved  quanta, 
as that of a hadron gas  and index the variables with subscript $_H$.
To phenomenologically describe a new phase, which includes
constituent quarks and anti-quarks (Q), characterized by
their mass and  by a vacuum pressure,
we generalize the equations of state to
\bea
\varepsilon_Q & = & m_Q n_Q + \lambda_\varepsilon n_Q T + B, \label{e:eos1m}\\
	p_Q   & = & \lambda_p n_Q T - B, \label{e:eos2m} 
\eea
where $\lambda_\varepsilon$ and $\lambda_p$ are constants.
Let us introduce the notation $\kappa_Q = \lambda_\varepsilon/\lambda_p$.
With this simple ansatz, 
the form of the relativistic Euler and temperature equations
does not change and we obtain the same form of the solutions for 
$n$, $T$ and ${\bf v}$ as before, however, with modified boundary
conditions corresponding to modified constants of integration.
We may refer to this new phase as a kind of quark matter, 
providing  simple model equations of state
for massive quarks, and a bag constant
corresponding to color deconfinement.
The gluons are assumed to be integrated out, providing mass
for the quarks. Such a picture is qualitatively 
supported by phenomenological
fits to the lattice QCD equation of state with quarks that pick up
the constitutent mass around $T_c$ and gluons that carry big
effective mass in the same temperature domain~\cite{levai-heinz}.

Using the above ansatz for the equations of state, we obtain
the following solution for the pressure:
\be
	p_Q = (p_{0,Q} + B) 
	\left( \frac{\tau_0}{\tau} \right)^{3 + 3/\kappa_Q} - B.
\ee
If we chose $B>0$ and $\lambda_p > 1$, then at low temperatures
the state consisting of non-relativistic massive ideal gas will be
stable, and at some critical temperature $T_c$ the pressure
of the two phases is balanced. However, at high temperatures,
the pressure of the new phase will be larger than that of the
hadronic matter, as $\lambda_p > 1$, hence at high temperatures
this phase will be the stable one. 

 The critical temperature, $T_c$ is defined by
\be
	n_H T_c = \lambda_p n_Q T_c - B.
\ee
For the hydrodynamical description, the specification of $n_H$ and
$n_Q$ as a function of e.g. a temperature and chemical potential is
not necessary, any thermodinamically consistent form can be included
into the hydro solutions that we focus on. Hence the above equation
may indicate, if $n_H$ and $n_Q$ are expressed as a function of 
the temperature and the (baryo)chemical potential, that the 
critical temperature may become density dependent.

One can show that the expansion is adiabatic if the matter consists of only
one of the phases. We may assume, that the expansion is adiabatic also
during the period of phase coexistence, and describe the phase transition
with the help of a Maxwell construction. This case is similar to the
construction discussed by Gyulassy and Matsui in case of a
massless ideal gas~\cite{gyu-matsui}. 
Let us denote the entrophy density of the two phases  
by $s_Q$ and $s_H$, and the volume fraction of the two phases by
$f_Q$ and $f_H = 1 - f_Q$, respectively.
The proper-time dependence of the entrophy density is then
given by
\be
	s(\tau)  = f_Q(\tau) s_{Q,c} + (1 - f_Q(\tau)) s_{H,c}.
\ee
At the beginning of the phase transition,  at $\tau= \tau_{Q,c}$
the whole system consists of ``Quark Matter", that is $f_Q(\tau_{Q,c}) = 1$.
By the time the phase transition ends, one obtains $f_Q(\tau_{H,c}) = 0 $.
As the well known steps of the Maxwell construction can be
copied from the case without conserved charges, for further
details we simply refer to section 6.2.3 of ref.~\cite{csernai}.

Starting the hydrodynamical expansion from  
 a quark matter initial state, 
the form of our hydrodynamical solution becomes
unmodified even during the phase transition, and  the coasting type
of hydrodynamical evolution can be continued at a fixed $T_c$,
from $\tau_{Q,c}$ to $\tau_{H,c}$ untill all matter is
transformed to hadronic matter. Then one can continue the
same kind of coasting solution untill the hadronic matter
freezes out. 

This expansion takes usually a long time, which is difficult to
justify by experimental evidences. However, the same kind of hydrodynamical
solution can be matched with a non-equilibrium type of phase transition,
the sudden rehadronization and a simultaneous freeze-out of hadrons
from a supercooled Quark Matter, as described in greater details
in refs.~\cite{csorgo-csernai,csernai-mishustin}. Then in the same model
one may reach a deeply supercooled Quark Matter state, and the 
transition to the hadron gas may proceed via a mechanical instability,
associated with a negative pressure state, governed by the conservation
of matter, energy and momentum and the impossibility 
of an entrophy decrease in a deflagration through 
a hypersurface with a time-like normal vector at
$\tau = \tau_{TD}$,
\bea
	\left[ T^{\mu \nu} df  n_\nu \right] & = & 0,\\
	\left[ n u^{\nu} df n_\nu \right] & = & 0,\\
	\left[ s u^{\nu} df n_\nu \right] & \ge & 0,
\eea
where $df n_\nu \equiv d\sigma_\nu$
is the normal-vector of the $\tau=\tau_{TD}$ hypersurface.
These equations and inequality can be solved very similarly to 
ref.~\cite{csorgo-csernai}, that discussed the case of
a transition from a massless ideal quark-gluon plasma (QGP)
to a massless hadron gas. The lack of acceleration,
and the invariance of the equations of state for rescaling 
the temperature and for adding a vacuum energy term are the
essential reasons why earlier considerations
that discussed phase transitions for massless particles
can be straigthforwardly implemented to our new
family of exact hydrodynamical solutions that describe the
phase transitions and expansions of massive quanta. 

 Recently, a new method has been proposed to evaluate particle
emission from hydrodynamically evolving, locally thermalized 
sources, that generalizes the Cooper-Frye freeze-out conditions
for systems with volume emission~\cite{sinyukov-hama,cont-em}. 
This method, based on the method of escaping probabilities,
can also be applied to evaluate the particle spectra and correlation
functions from the new hydrodynamical solutions, as the 
non-relativistic form of the solutions corresponds to the
limiting behaviour that was discussed in ref.~\cite{sinyukov-hama}.

\section{Factorized solutions}
From the structure of the new family of solutions presented above, 
it is obvious how to generate further new solutions of 
relativistic hydrodynamics. 

Observe that a modified version of the scaling variable of 
eq.~(\ref{e:ssol}) can be defined, and another family of the 
same kind of ellipsoidal solutions can be generated by introducing 

\bea
 s^\prime   & = & 
	\frac{r_x^2}{\dot X^2_0 \tau^2} +
	\frac{r_y^2}{\dot Y^2_0 \tau^2} + 
      \frac{r_z^2}{\dot Z^2_0  \tau^2}\,, \label{e:spsol}\\
	X^\prime & = & \dot X_0 \tau , \label{e:xsolp}\\
	Y^\prime & = & \dot Y_0 \tau , \label{e:ysolp}\\
	Z^\prime & = & \dot Z_0 \tau , \label{e:zsolp}
\eea
where the dots should be understood as derivatives with respect to 
$\tau$. Mathematically, the modified scaling variable solves the 
relativistic hydro equations, because it also satisfies the 
requirement
\begin{equation} 
	u^\mu \partial_\mu s =
	u^\mu \partial_\mu s^\prime\, =\, 0\,, \label{e:sdif}
\end{equation} 
if the flow is scaling. This condition can be considered also as the 
criterium of a ``good" scaling variable, and satisfied by the 
following independent sets of variables: 
\bea
	s_x & = & \frac{r_x^2}{t^2}\,, \\
	s_y & = & \frac{r_y^2}{t^2}\,, \\
	s_z & = & \frac{r_z^2}{t^2}\,,
\eea
or alternatively, 
\bea
	s_x^\prime & = & \frac{r_x^2}{\tau^2}\,, \\
	s_y^\prime & = & \frac{r_y^2}{\tau^2}\,, \\
	s_z^\prime & = & \frac{r_z^2}{\tau^2}\,.
\eea
All of the 6 scale variables defined above satisfy eq.~(\ref{e:sdif}) 
if the flow profile is of a 3-dimensional scaling type, 
$u^\mu = x^\mu/\tau\,$.

As the relationship $\NN(s) = 1/\TT(s)$ holds in the class of 
solutions we present here, if we assume a factorized form for the 
scaling function of the density, we automatically generate a 
factorized form for the scaling function of the temperature, hence it 
becomes easy to generate further new solutions. Let us define the 
scaling functions $\NN_x(s_x)$, $\NN_y(s_y)$ and $\NN_z(s_z)$ that 
are positive, differentiable and satisfying 
$\NN_x(0) = \NN_y(0) = \NN_z(0) = 1$, otherwise being abritrary and 
independent from each other. Then, a new type of hydro solutions 
reads 
\bea
 {\mathbf v} & = & \frac{\mathbf r}{t}\hspace{2.cm}{\rm or}  
       \hspace{2.cm}u^\mu=\frac{x^\mu}{\tau}\,, \\
 n(t,{\mathbf r}) &=& n_0 \left(\frac{\tau_0}{\tau}\right)^3 
	 \NN_x(s_x) \NN_y(s_y) \NN_z(s_z)\,, \label{e:nf} \\
 (t,{\mathbf r}) &=& p_0 \left(\frac{\tau_0}{\tau} 
       \right)^{3 +  3/\kappa}\,, \\
 T(t,{\mathbf r}) &=& T_0 \left(\frac{\tau_0}{\tau}\right)^{3/\kappa} 
		\frac{1}{\NN_x (s_x)}
		\frac{1}{\NN_y (s_y)}
		\frac{1}{\NN_z (s_z)}\,. \label{e:tf} 
\eea
Note that this form of solution is invariant for a change of the 
scaling variables and the scales as 
$(s_x, s_y, s_z)\rightarrow(s_x^\prime, s_y^\prime, s_z^\prime)$ and 
$(X,Y,Z)\rightarrow(X^\prime, Y^\prime, Z^\prime)$, with the time 
derivatives in the definitions of the Hubble flow field, 
eqs.~(\ref{e:vx}-\ref{e:vz}) understood as derivations with respect 
to $\tau$ as in eqs. (\ref{e:spsol}-\ref{e:zsolp}). 

\section{General form of solutions, beyond ellipsoidal symmetry}
The key point in checking that the above factorized forms indeed 
solve the equations of relativistic hydrodynamics is that the 
comoving derivatives of the scaling functions, 
$u_\mu \partial^\mu \NN_i(s_i)$, are proportional to 
$u^\mu \partial_\mu s_i\,$, hence vanish. This is, together with 
$u^\mu\partial_\mu u_\nu =0$, the essential property of the scaling solutions. 

We find that the most general form of the scaling variable is
\begin{equation} 
\overline{s} = F(s_x,s_y,s_z) \equiv G(s_x^\prime,s_y^\prime,s_z^\prime)\,,
\end{equation} 
which means that any function of scaling variables $(s_x, s_y, s_z)$ 
or $(s_x^\prime, s_y^\prime, s_z^\prime)$ can be utilized as a new 
scaling variable. Indeed, we have
\begin{equation} 
u_\mu \partial^\mu F(s_x,s_y,s_z) = 
		\sum_{i=x,y,z}\frac{\partial F}{\partial s_i}  \, 
		u_\mu \partial^\mu s_i = 0\,.
\end{equation} 
which yields the generalized form of the new family of solutions 
of relativistic hydrodynamics:
\bea
{\mathbf v} & = & \frac{\mathbf r}{t}\hspace{2.cm}{\rm or}  
        \hspace{2.cm}u^\mu=\frac{x^\mu}{\tau}\,, \\
n(t,{\mathbf r}) &=& n_0\left(\frac{\tau_0}{\tau}\right)^3 
        \NN(\overline{s})\,, \label{e:no} \\
p(t,{\mathbf r}) &=& p_0 \left(\frac{\tau_0}{\tau}
                         \right)^{3 + 3/\kappa}, \\
T(t,{\mathbf r}) &=& T_0 \left(\frac{\tau_0}{\tau}\right)^{3/\kappa}
                \frac{1}{\NN(\overline{s})}\,,
			\label{e:to}
\eea
with the constraint that $\NN(0) = 1$. 
For example, if we choose $\overline{s} = s_x + s_y + s_z\,$, we 
obtain solutions with ellipsoidal symmetry, however, one may choose 
$\overline{s} = s_x + s_y - s_z$ to obtain solutions where the 
isotherms are one-sheeted hyperboloids, or 
$\overline{s} = - s_x - s_y + s_z$ where the temperature and the 
density are constants on two-sheeted hyperboloids. In fact the 
possibilities are infinitely rich.

\section{Non-relativistic limiting behaviour}
As the considered equation of state contains mass as a free 
parameter, it is possible to study the non-relativistic limiting 
case of these new relativistic solutions. The local thermal motion 
is non-relativistic if $m \gg T$, the flow is non-relativistic in 
the region of $|{\bf r}| \ll t$, which implies $\tau \approx t$ and 
relates the relativistic solutions with ellipsoidal symmetry to the 
asymptotic, large time behaviour of the ellipsoidally symmetric, 
non-relativistic solutions of refs.~\cite{ellsol,ellsp,csel}. 
Let us recapitulate here the general form of the non-relativistic 
solutions of fireball hydrodynamics~\cite{csel}.

The definition of the scaling variable $s$ coincides with that of 
eq.~(\ref{e:sdef}). The solution for the flow field coincides with
that of eqs.~(\ref{e:vx}-\ref{e:vz}). The scaling function for the
temperature, $\TT(s)$ is utilized to express the solution for the
NR density and temperature fields as 
\bea
 n & = & n_0\frac{X_0 Y_0 Z_0}{XYZ}\frac{1}{\TT(s)} 
      \exp\left[-\frac{m(\dot X^2_{\rm as}+\dot Y^2_{\rm as} 
      +\dot Z^2_{\rm as })}{T_0}
	\int_0^s du \frac{1}{\TT(u)}\right], \\
 T & = & T_0\left(\frac{X_0 Y_0 Z_0}{XYZ}\right)^{1/\kappa}\TT(s)\,. 
\eea 
The large time behaviour of these non-relativistic, self-similar, 
ellipsoidal solutions of hydrodynamics is characterized by constant 
values of 
$(\dot X_{\rm as}, \dot Y_{\rm as}, \dot Z_{\rm as})$, 
which implies that for asymptotically long times a scaling flow field 
develops also in these non-relativistic solutions, 
${\mathbf v} \approx {\mathbf r}/{t}$. 
In the exponential factor that appears in this non-relativisitic 
solution for $n$ we may expand the inverse scaling function 
$1/\TT(u)$ as a polinomial in $u$. Keeping the leading order terms, 
and performing the integration, we find that the exponential factor 
will be approximately a Gaussian factor. However, the widths of the 
Gaussians in all the directions will be proportional to the time. As 
we have assumed that $|{\bf r}| \ll t$, this exponential factor 
yields a factor of $1 + {\mathcal O}({\bf r}^2/t^2)$, hence we find 
the general asymptotic behaviour of the non-relativistic solutions to 
be the {\it same form } as the general solution of the 
{\it relativistic} solution of the hydrodynamical equations: 
\bea
 n_{\rm as}&\approx& n_{\rm as}\frac{t_0^3}{t^3}\frac{1}{\TT(s)}\,
   ,\\ 
 T_{\rm as}&\approx& T_{\rm as}\left(\frac{t_0}{t}\right)^{3/\kappa} 
   \TT(s)\,.
\eea
Due to the rescaling, the constants of normalization $n_{\rm as}$
and $T_{\rm as}$ are different from $n_0$ and $T_0$.

\section{Summary} We have generalized the recently found new family 
of solutions of relativistic hydrodynamics to the case of expanding 
fireballs with ellipsoidal symmetry. The solutions contain an 
arbitrary scaling function $\NN(s)$, restricted only by 
non-negativity and by the requirement of $\NN(0) = 1$, a very rich 
set of possible scaling variabes $s$, and 5 important parameters, 
the mass $m$, the parameter $\kappa$ of the equation of state, the 
scale parameters $\dot X_0$, $\dot Y_0$ and $\dot Z_0$.
Furthermore, we generalized the equations of state to describe
a phase transition from a deconfined state consisting of massive 
(constituent) quarks and antiquarks to a state consisting of massive
hadrons. Based on the scaling properties of the equations of state,
as well as on the coasting nature of the expansion, we have shown
that both the usual adiabatic Maxwell construction as well as the
fast, non-equilibrium time-like deflagrations can be constructed
and described within the considered class of hydrodynamical solutions.

We expect that some of these solutions of relativistic
hydrodynamics may have future applications 
in non-central collisions of heavy ions at relativistic bombarding 
energies, a topic of great current research interest.

\section*{Acknowledgments}  
T. Cs. would like to Y. Hama and G. Krein for kind hospitality 
during his stay at USP and IFT, Sao Paulo, Brazil, and M. Gyulassy 
for his kind hospitality at Columbia University, New York. This work has 
been supported by a NATO Science Fellowship (Cs.T.), the OTKA 
grants T026435, T029158 and T034269 of Hungary, the NWO - OTKA 
grant N 25487 of The Netherlands and Hungary, the FAPESP grants 
99/09113-3 00/04422-7 and 02/11344-8 of S\~ao Paulo, Brazil,
and by the US DOE grants DE-FG02-93ER40764 and DE-FG02-01ER41190.

\vfill\eject
\end{document}